\documentclass{article}
\usepackage{bm}
\usepackage{hyperref}
\usepackage{float}
\usepackage[utf8]{inputenc}
\usepackage{braket}
\usepackage{url}
\usepackage{amsfonts}
\usepackage{tabularx}
\usepackage{bbold}
\usepackage{graphicx}
\usepackage{caption}
\usepackage{subcaption}
\usepackage{appendix}
\usepackage{comment}
\usepackage{float}
\usepackage{amsmath}
\usepackage{tabularx}
\usepackage{biblatex}
\bibliography{references.bib}
\graphicspath{ {images/} }

\title{A Review of Quantum communication using high-dimensional Hilbert spaces}
\author{Yuval Idan and Avihai Didi }

\begin{document}

\maketitle

\begin{abstract}
In this project we examine several different quantum key distribution protocols which we divide into ones utilizing qubits whose Hilbert spaces are two dimensional and ones whose Hilbert space dimension is greater than two, these units of data in quantum computers are known as qudits and in the papers we'll examine are implemented using the orbital angular momentum of twisted photons.
In sections \ref{sectionqubitprotols} and \ref{sectionquditprotocols} the specific procedures of each protocol are briefly described and followed by an examination of the theoretical and experimental merits of each protocol.
These merits are measured in the bit error rate tolerance $e_b$, which quantifies the maximum channel noise and the key rate $R$ which quantifies the rate at which data is transferred in the protocol.
In section \ref{sectionFinalComparison} we present a unified view of all the relevant data for the different protocols, and argue for the benefits and drawbacks of the different protocols for different applications.    
\end{abstract}

\section{Introduction}
In classical computers the basic unit of information is a bit.
Each bit can equal $0$ or $1$, and using a combination of multiple bits we are able to encode any data we should wish using $\log_2 (|\Sigma|)$ bits where $\Sigma$ is the alphabet describing the data in question and $|\Sigma|$ is the cardinality (size) of the alphabet.
Different types of data are represented by different alphabets, such as unicode for text, RGB values for images, and so on.
In principle, the use of a binary unit of information is arbitrary, and units of information that can take on more than two values could equivalently be used, in which case the number of such units needed to represent information in some alphabet $\Sigma$ will be $\log_d (|\Sigma|)$ where $d$ is the number of different values the basic unit of information can take on.
However, in practice testing has definitively shown that using a binary unit of data is much simpler and cheaper to implement, and as such it is the dominant form of data storage in classical computers

In quantum computing, the basic unit of information is the qubit.
Qubits can take on either of the values a classical bit may take on, but may also be in a state of super-position of the two, in practice these qubits are usually implemented using two level systems such as the spin of a spin-half particle, a super conduction circuit, or the polarization of a photon.
Said super-positions may be manipulated using methods studied within the field of quantum mechanics, and this manipulation forms the basis of the majority of quantum computing.
Qubit are commonly represented using the Bloch sphere, which, for a qubit in a normalized superposition of the states $\ket{0}$ and $\ket{1}$ with a phase difference of $\phi$ is written as
\begin{equation}
    \ket{\psi} = \cos{(\theta/2)} \ket{0} + e^{i \phi} \sin{(\theta /2)} \ket{1}
\end{equation}
This representation gives an intuitive visualization method for the effects that various quantum operators have when acting on a qubit.

\begin{figure}[H]
  \centering
  \includegraphics[width=0.8\textwidth]{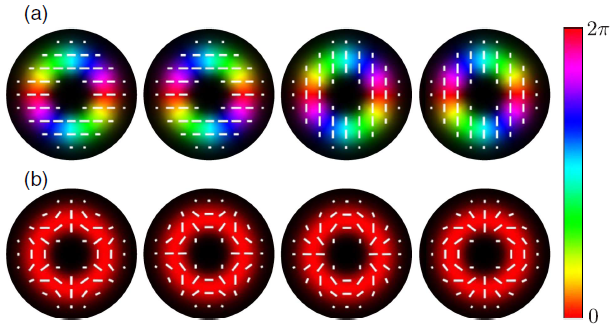}
  \caption{(Image taken from \cite{Sit17}).
  A pair of mutually unbiased bases for $l = 2$, encoding information in OAM (a) and polarization (b) where the intensity is used to denote the amplitude, the white lines denote the transverse polarization, and the color is used to denote the phase of the wave function.
  Each basis is orthonomalized and mutually unbiased with respect to the other basis such that the absolute value squared of the projection of any state from one basis onto the other equals $1/4$.
  These MUBs posses identical intensity profiles and are shape-invariant upon free space propagation, making them suitable for long range communication.
  }
  \label{figtwistedphotons}
\end{figure}

Analogously to classical computing, information in quantum computers may also be represented by units whose Hilbert spaces have a dimension greater than two, and these units are called qudits.
Just like qubits, there are many competing ways to implement qudits, although no specific implementation has been definitively shown to be the best means of implementation yet (in the way that the voltage drop on a circuit has for classical bits) photon based solutions seem promising due to their long decoherence time and comparatively weak interactions with other particles.
In addition to polarization which is commonly used as a binary unit of data in qubit based systems, a photon may also carry orbital angular momentum (OAM) \cite{ogOAMpaper}, corresponding to helical wavefronts.
Whereas polarization is naturally bi-dimensional, being typically represented using the basis $\{ \{ \ket{L} \text{, } \ket{R} \} \text{, } \{ \ket{\leftrightarrow} \text{, } \ket{\updownarrow} \} \} \}$ (or any other complete $2D$ basis), the dimensionality of OAM by comparison is infinite ($\aleph_0$ to be exact), as it can take on any value of $\hbar l$ where $l$ is an integer \cite{naturepaperaboutOAM, Sit17}.
These light waves which are in an arbitrary coherent superposition of different polarization of and spatial modes are referred to as structured photons, a visual example of some of the states that the photons can take on is given in figure \ref{figtwistedphotons} ,and in the examples we'll examine later in this project the OAM of these light waves is used to experimentally implement qudits.

While the practicality of using units of information which can take on more than two states has been thoroughly ruled out by now in the field of classical computation, the same can not be said for the emerging field of quantum computation, where the applicability of such units of data to real world use cases remains up for debate.
This possibility is the primary focus of our work here, we will study it by comparing the application of qubit and qudit based protocols for quantum key distribution (henceforth QKD), to see the advantages and disadvantaged of qudits compared to qubits.
It should also be noted that unlike in classical computers where the only noteworthy application of units of information that can take on more than two states is the possibility of representing data using less than $\log_2 (|\Sigma|)$ units of information, as we'll see in the protocols we'll examine the higher dimension can instead be used to improve the error resilience of the protocol or decrease the probability that an eavesdropper could successfully intercept vital information while still only using the qudit to represent one classical bit of data.

Our primary goal in this project is to form a comprehensive overview of the current state of research, theoretical and experimental alike, into the applicability of qudit based protocols for quantum key distribution and to contrast those with analogous qubit based protocol so as to ascertain the relative benefits and drawbacks of each.
It should be noted that the applicability of qubit based solutions to problems besides QKD lies outside the scope of this project and should be considered separately.

\section{Quantum Key Distribution (QKD)}
In order for two computers to safely communicate without fear that their data is being read by a third party it is common practice for them two to encrypt the information so that even if a third party does intercept the data being sent they would have no use for it without the decryption key.
The issue then becomes how the two could share an encryption key between them over a compromised channel, since if they simply send it as is the key itself could also be intercepted.
This issue was solved using the RSA algorithm, first published in 1977.
The algorithm relies of the fact that the problem of integer factorization is an NP problem, meaning that unless $P=NP$ is proven (which most mathematicians think is highly unlikely) RSA based encryption remains impractically difficult to solve on a classical computer, making any data encrypted with it, while technically possible to decrypt, completely safe from the eyes of third parties, as the average time it would take to decrypt any data encrypted using RSA is many orders of magnitude greater than the age of the universe.

While RSA encryption is practically impregnable by classical computers, in 1994 Peter Shor published an algorithm designed for quantum computers which can solve integer factorization problems in polynomial time, which in theory allows it to crack RSA encryption with relative ease.
The development of Shor's algorithm greatly endangered the safety of classical key distribution and necessitated a transition to quantum based key distribution methods.
Thankfully by this point researches have already began developing such methods, which, once implemented, allow for the sharing of encryption keys without risk of Shor's algorithm.
At time of writing, current quantum computers are not capable of running Shor's algorithm for numbers large enough for them to pose an actual threat to RSA encryption, but the study of QKD protocols seeks to find a solution to this problem before it manifests.

\section{Qubit based protocols}
\label{sectionqubitprotols}
\subsection{BB84 protocol}
\label{secQuBitBB84}
The BB84 protocol operates using qubits which are represented using the orthonormalized wave functions $\ket{0}$ and $\ket{1}$ .
\subsubsection{Description of the protocol}
In the first step of the protocol, Alice, who wishes to send an encryption key to Bob, generates an $n$ bit long key which we'll label $\Vec{a}$ and a second $n$ bit long vector which we'll call $\Vec{b}$, both should be chosen randomly.
She then constructs the wave function that she'll send to Bob such that each of the $n$ qubits is determined by both $\Vec{a}$ and $\Vec{b}$ as follows:
\begin{equation}
\begin{split}
    \ket{\psi_{00}} &= \ket{0}
    \\
    \ket{\psi_{10}} &= \ket{1}
    \\
    \ket{\psi_{01}} &= \ket{+} = \frac{1}{\sqrt{2}} (\ket{0} + \ket{1})
    \\
    \ket{\psi_{11}} &= \ket{-} = \frac{1}{\sqrt{2}} (\ket{0} - \ket{1})
\end{split}
\end{equation}

\begin{figure}[H]
  \centering
  \includegraphics[width=0.5\textwidth]{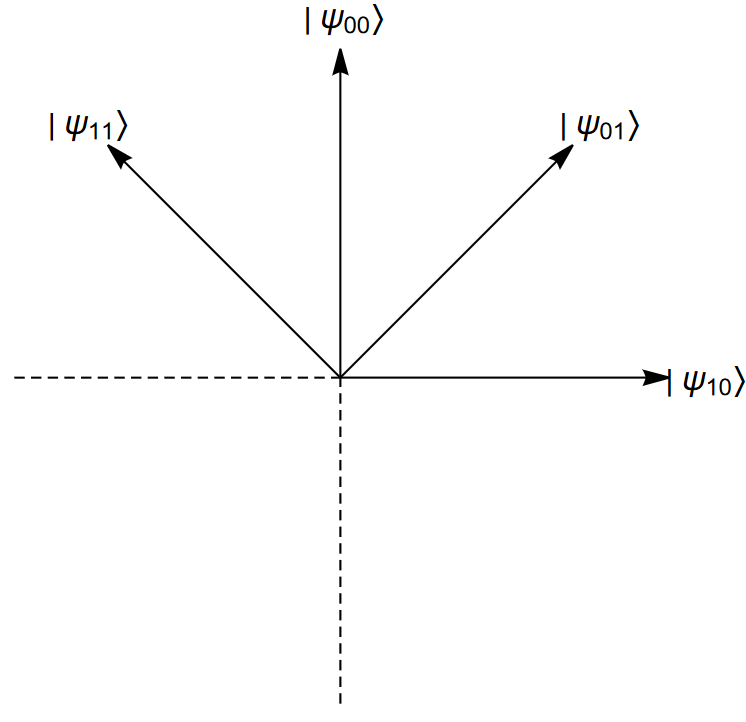}
  \caption{A schematic representation of the four states used in the BB84 protocol. The squared projection of any state with another of it's basis is zero whereas the squared projected amplitude onto a vector from the other basis is $1/2$.}
\end{figure}

Next, Alice sends this wave function over to Bob over a public channel, it is assumed that Eve has access to this channel and can freely measure the wave function Alice intended to send to Bob and then resend the resulting wave function to him.
Once Bob receives the wave function, he generates his own $n$ bit long vector which we'll call $\Vec{c}$, this vector determines the basis which he'll use when measuring each of the qubits in the wave function he received, where for $0$ he'll measure in the $\{ \ket{0} \text{, } \ket{1} \}$ basis and for $1$ he'll measure in the $\{ \ket{+} \text{, } \ket{-} \}$ basis.
Bob will save the results of this measurement as the "Raw Key".
When measuring in the same basis that Alice happened to choose, which on average will happen for $n/2$ if the qubits, Bob will receive the exact bit that Alice intended to encode, whereas for the cases where he chooses the wrong basis the result he'll get is random.
After this, Alice publicly announces the basis she encoded her key in $\Vec{b}$ over a classical communication channel.
Hearing this, Bob will keep only the parts of the raw key where his basis of measurement $\Vec{c}$ happened to match Alice's basis.
Assuming that no noise or third party was involved, the results of his measurements there will allow him to recover the a key which will on average be $n/2$ bits long.
If Alice or Bob suspect that there may have been eavesdropping they can publicly announce the results of a subset of the key bits so as to compare what they got, if no eavesdropping attempt was made the two of them should obtain the same result but if Eve interfered with the measurement the two subsets will be different and Alice and Bob will know that they should discard the key and try again.

\subsubsection{Key rate and error resistance}
The key rate for the BB84 protocol in 2 dimensions is:
\begin{equation}
    R=1-2h(e_b)
\end{equation}
where $h(x)=-x \log_2(x)-(1-x) \log(1-x)$ is the Shannon entropy $e_b$ is the QBER.
The key rate is a measure of the average number of bits of information that Bob will be able to recover from each unit of information Alice sends to him which has passed trough the sifting process, those units being single qubits for qubit based protocols and single qudits for qudit based protocols.
It should be noted that this is not necessarily the same as the average number of bits Bob is able to recover from each unit of information Alice sends, that would be given by the product of the key rate and the sifting rate which is always less than one.
We address this distinction and its implications in greater detail in section \ref{sectionFinalComparison}.
The QBER is defined as the probability that a generic bit in Bob’s sifted
string is different from the corresponding bit in Alice’s sifted string \cite{Pirandola20aaaaaaaaaaaa}.
In order to compute the QBER, Alice and Bob perform a session of parameter estimation, where they agree to disclose a random subset of their data. Comparing these bits (later discarded), they can quantify the QBER and check if this is lower or higher than a certain security threshold of the protocol.

The quantum bit error rate (henceforth QBER) threshold for the BB84 protocol is $e_b = 11\%$, meaning that the protocol is only reliable so long as the channel the information is being transmitted on only flips bits less than $11 \%$ of the time.
One way to increase the QBER threshold is the six state protocol.
The six state protocol is the brother of BB84, but with all the three MUB in two dimension. 
Thus, $\ket{x},\ket{y},\ket{z}$, because there are more options to Eve she produce more QBER then the normal BB84, In addition the six state protocol have more noise tolerance then the BB84.
The unconditional key rate against coherent attacks has the following expression \cite{Pirandola20aaaaaaaaaaaa}:
\begin{equation}
    R_{\text{6state}}=\frac{1}{3}(1+\frac{3e_b}{2})\log_2(\frac{e_b}{2})+(1-\frac{3e_b}{2}) \log_2(1-\frac{3e_b}{2}))
\end{equation}
The security threshold value is about $12.6\%$, slightly improving of the BB84 protocol and have more immunity against eavesdrop.
The security threshold is a measure of how much noise Eve is able to make trough her eavesdropping attempts before she is detected relative to the allowed QBER.
This is the MUB version of the 2d BB84 protocol.
We discuss higher dimensional MUB protocols in section \ref{sectionquditMUB}.

\subsubsection{Eavesdropping immunity}
In quantum information the medium we use for communication is noisy, and this noise is ascribed to Eve the Eavesdropper in the worst-case scenario.
Eve will attempt to intercept the communications send to Bob, measure them herself, and then resend the post-measurement result to Bob.
The issue Eve faces is that she has a $50 \%$ chance of picking the wrong basis for her measurement, for each qubit.
In the cases where she picks wrong, even if Bob happened to pick the same basis as Alice the results he'll obtain will be random due to the collapse of the wave function which occurred when Eve measured it.
The probability that they all happen to pick the same basis and Eve's eavesdropping goes unnoticed is $1/4$ per qubit
The total probability that Eve's eavesdropping won't be detected is $(1/4)^n$ which exponentially converges to zero, meaning that the protocol is safe from eavesdropping.
This safety from eavesdropping is guaranteed by the fact that any attempt at intercepting the message Alice sent to Bob by Eve will create a lot of noise in the system which will bring the QBER up to $e_b = 25\%$ \cite{Pirandola20aaaaaaaaaaaa} significantly higher than the allowed rate of $11\%$.
In practice Eve can also use the faults of the system (false positives/negatives in the detector, imperfections in the pulses generated by the laser, etc...) in order to gain information about the key which is independent of the protocol.
In section \ref{sectionQuantumCloningMachines} we discuss one example of this which is the use of high dimensional quantum cloning machines, which can produce imperfect copies of the data Alice sent, to intercept the information.

\subsection{Chau02 protocol}
This protocol was introduced by H. F. Chau in this 2002 paper \cite{chau02}.
We've taken to calling it Chau02 after we were unable to find a specific name for it in the literature.
It is the qubit based QKD protocol with the highest bit error rate tolerance according to \cite{chau15OGpaper, chau02} with a tolerance of up to $27.6 \%$.
In the introduction of \cite{chau15OGpaper} H. F. Chau directly contrasts the bit error rate tolerance of this protocol with that of the qudit based version, we will address this comparison in greater detail in section \ref{sectionFinalComparison} after we introduce the higher dimension version of this protocol in section \ref{subsectionChau15}.

\subsubsection{Description of the protocol}
Alice generates a set of $n$ qubits each of which is randomly set to one of the six eigenstates of the Pauli matrices $\{ \sigma_x \text{, } \sigma_y \text{, } \sigma_z \}$.
The total wave function of the $n$ qubits $\ket{\psi}$ is then sent to Bob over an insecure channel.
Bob acknowledges the reception of the qubits and then randomly chooses one of the three Pauli matrices for each qubit, he then measures each qubit in the eigenbasis of the Pauli matrix he chose for it.
Once this step is complete, Alice and Bob both publicly announce the basis they used over a classical channel and keep only the ones they both happened to prepare in the same basis.
The next (several) steps of this protocol consist of the application of several privacy amplification procedures which lie outside the scope of our word, suffice it to say that this process greatly improves the maximum bit error rate the protocol can handle.

\subsubsection{Key rate and error resistance}
The theoretical bit error rate for this protocol is $e_b^{\text{max}} = 1/2 - 0.1 \sqrt{5} \approx 27.6 \%$, this gives us a theoretical key rate of $R(0) = 1$.
We were unfortunately unable to find any papers detailing an experimental implementation of this protocol, so the theoretical maximum values are all we have to work with.

\section{Qudit based protocol(s).}
\label{sectionquditprotocols}
\subsection{BB84}
\label{sectionQuditBB84}
Following up on our previous discussion of BB84 in section \ref{secQuBitBB84} where we presented the 2 dimensional version of the protocol, we'll now show the $d$ dimensional version.
The primary advantage of increasing the dimension of the data carrying unit we use by switching from qubits to qudits is that the number of mutually unbiased basis (MUBs) with which Alice may encode information increases.
In $d$ dimensions we have $d+1$ MUBs where $d$ is a power of a prime number, for instance, for $d=2$ we can use the eigenbasis of the three Pauli matrices as a set of 3 MUBs.
In the paper we use as a source for the discussion of this protocol \cite{Bouchard2018experimental} high dimensional BB84 is implemented in $d = 2, 4, 8$ dimensions.
As the first basis we use pure orbital angular momentum (OAM), i.e $\ket{\phi} \in \{-d/2,...,d/2 \}$.
For the second basis we use the Fourier basis which is given by the discrete quantum Fourier transform of the position space basis $\ket{\phi_i}=\frac{1}{\sqrt d}\sum_{j=0}^{d-1}w_{d}^{ij}\ket{\psi_i}$ with $w_d=e^{2\pi i/d}$.
The explicit form of this basis is given in \cite{explicitFormOfTheQFTbasis}.
\subsubsection{Key rate and error resistance}
The experimental quantum bit error rate results obtained in \cite{Bouchard2018experimental} for the three dimensions tested are $e^{d=2} = 0.628\%$, $e^{d=4} = 3.51 \%$, and  $e^{d=8} = 10.9\%$.
The experimental quantum key rate results obtained are $R^{d=2} = 0.8901$, $R^{d=4} = 1.4500$, and $R^{d=8}= 1.3942$.
Although the maximum key rate found in this paper is at $d = 4$, the higher bit error tolerance of $d = 8$ might make it preferable for real life applications for the particular implementation used here (Orbital angular momentum of twisted photons).

\subsection{Chau15 protocol}
\label{subsectionChau15}
The Chau15 protocol, described in \cite{chau15OGpaper} uses $2^N$ dimensional qudits, where $N$ is an integer greater than one, which were experimentally implemented as the orbital angular momentum of photons in \cite{Bouchard2018experimental}.
The basis vectors of the Hilbert space are $\cup_{i=1}^{2^N} \{ \ket{i} \}$.
\subsubsection{Description of the protocol}
First, Alice generates a string of $n$ bits which we'll denote as $\Vec{a}$.
She then generates a wave-function consisting of $n$ qudits as follows.
For each bit $a_k$ in $\Vec{a}$, Alice chooses a random pair of numbers $\{ i \text{, } j \}$ where $i \neq j$ and encodes the bit as $(\ket{i} + (-1)^{a_k} \ket{j}) / \sqrt{2}$.
Once this is done for every bit in $\Vec{a}$ Alice transfers the completed wave function to Bob, the medium used to transfer the data is presumed to be unsafe.
Once Bob has received the wave function, for each qudit he chooses a random pair of numbers $\{ i' \text{, } j' \}$ where $i' \neq j'$ and measures the qudit in the subspace $\{ (\ket{i'} + \ket{j'}) / \sqrt{2} \text{, } (\ket{i'} - \ket{j'}) / \sqrt{2} \}$.
He keeps all of the $i' \text{, } j'$ pairs he kept and all of the results of the measurements. If a measurement of a subspace showed that the qudit is $(\ket{i'} + \ket{j'}) / \sqrt{2}$ he saves it as zero, and otherwise he saves it as one.
At this point, Alice publishes the list of $i \text{, } j$ pairs that she picked on a classical public channel, Bob in turn publishes his list of $i' \text{, } j'$ pairs.
The two of them then keep only the elements of the key where $i = i'$ and $j = j'$.
By checking if the results he obtained by measuring projecting these qudits onto $(\bra{i'} + \bra{j'})/\sqrt{2}$ is able to fully determine the elements of $\Vec{a}$ which correspond to these qudits.
Alice and Bob can now use those elements of $\Vec{a}$ as a key.
\subsubsection{Key rate and error resistance}
For $4 \leq N$ ($16 \leq d$) the theoretical bit error rate tolerance of the Chau15 protocol is $50 \%$ \cite{chau15OGpaper}.
In \cite{Bouchard2018experimental} the bit and dit error rates, as well as the key rates were tested for $N = 2$ and $N = 3$ ($d=4$ and $d=8$ respectively).
For $N=2$ ($d=4$) they obtained an average bit error rate of $e_b^{d=4} = 0.778 \%$, an average dit error rate of $e_d^{d=4} = 3.79 \%$ ,and an average key rate of $R^{(d=4)} = 0.8170$ bits per sifted photon.
For $N=3$ ($d=8$) they obtained an average bit error rate of $e_b^{d=8} = 3.11 \%$, an average dit error rate of $e_d^{d=8} = 0.82 \%$ ,and an average key rate of $R^{(d=8)} = 0.8172$ bits per sifted photon.
Theoretically, given a small enough dit error rate, bit error rates of up to $e_b^{max} = 50\%$ may be tolerated.
For orbital angular momentum of twisted photons which were the qudits used in this experiment this isn't a clear advantage since bit and dit error rates will usually correlate since they are the result of similar sources (e.g apparatus misalignment, turbulence, or optical aberration).
However, for other kinds of high dimensional quantum information carriers where the distinction between bit and dit error rates is less ambiguous the Chau15 protocol may be preferable \cite{Bouchard2018experimental}.

\begin{figure}[H]
  \centering
  \includegraphics[width=0.7\textwidth]{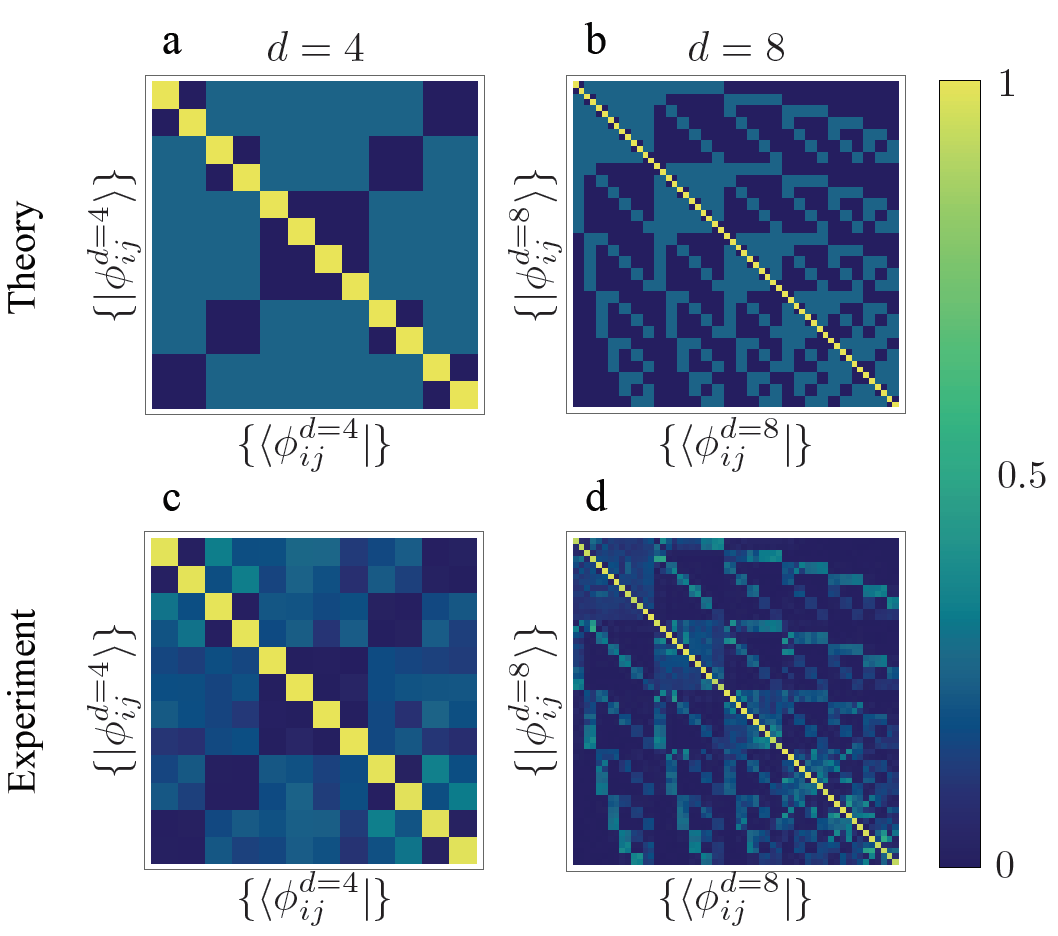}
  \caption{(Image taken from \cite{Bouchard2018experimental}).
  Theoretical predictions and experimental results for the probability of detection matrices for the Chau15 protocol in dimensions $d=4$ and $d=8$.
  The strong agreement between the theory and experiment is clearly visible.
  The rows correspond to the states sent by Alice whereas the columns correspond to Bob's projections of those states made by Bob during his measurements.
  The sifted data corresponds to the on-diagonal 2 by 2 blocks.
  }
\end{figure}

\subsection{MUB-based protocol}
\label{sectionquditMUB}
\subsubsection{Description of the protocol}
A MUB protocol is one wherein all $d+1$ MUBs that are available in dimension $d$ are used.
By using all of the MUBs we can perform QKD protocols (such as BB84) at a higher effective dimension than the one actually possessed by the qudits we use.
We can also only use some of the MUBs, choosing any amount between $2$ and $n+1$.
In some cases it might be beneficial to use less MUBs than we can since they have a scaling efficiency of $1/(1+d)$.
One example of an MUB protocol is the six state protocol in dimension two that we mentioned in section \ref{secQuBitBB84}, which uses all six eigenfunctions of the Pauli matrices, as opposed to just the eigenfunctions of $\sigma_x$ and $\sigma_z$.
The protocol may be generalized to high dimensions, and in \cite{Bouchard2018experimental} the two and four dimensional versions were considered.
\subsubsection{Key rate and error resistance}
For $d=2$, \cite{Bouchard2018experimental} obtained a QBER of $e_b^{d=2}=0.923\%$ and a key rate of $R^{d=2,m=3}=0.8727$.
For $d=4$, they obtained a QBER $e_b^{d=4}=3.87\%$ and a key rate of $R^{d=4,m=5}=1.5316$.

\section{High-dimensional quantum cloning}
\label{sectionQuantumCloningMachines}
A high dimensional cloning machine was proposed by \cite{Boucharde1601915}, although prefect cloning of an unknown quantum wave function is forbidden by the no cloning theorem, it nevertheless remains possible to create an imperfect clone of a wave function.
This was the subject of the paper at hand where they considered the potential venerability of high dimensional quantum states, such as qudits, to attempts to intercept and eavesdrop on communications without detection.
In the paper this issue was considered by first defining a variable called the cloning fidelity which was denoted as $F$.
This variable is the absolute value squared overlap of the imperfect clone onto the original wave function $F = |\braket{\psi_{\text{clone}}|\psi_{\text{original}}}|^2$, and by the no cloning theorem it is always smaller than one.
The optimal cloned state allowed by the rules of quantum mechanics has a bounded cloning fidelity of $F_{est}=\frac{2}{1+d}$, where d is the dimension of the quantum state.
Here, we concentrate on the $1\xrightarrow{}2$ universal optimal quantum cloning machine, for which the optimal fidelity of the two cloned copies is given by $F_{clo}=\frac{1}{2}+\frac{1}{1+d}$.

\subsection{Cloning attack}
As mentioned, the primary focus of the paper \cite{Boucharde1601915} was the potential application of such a cloning machine to eavesdrop on a quantum system, such as the transmission of an encryption key over a vulnerable network.
In the paper this was experimentally tested by using a cloning machine to attempt to attack a high dimensional version of the BB84 protocol (the one we described in section \ref{sectionQuditBB84}), the researches were able to demonstrate that the high dimension of the protocol made Eve's attempt's at interception clearly visible to Alice and Bob even without the utilization of any error correction or privacy amplification procedures, as Eve's attempted cloning attack brought the experimentally measured bit error rate up from $16 \%$ to a clearly noticeable $57 \%$, more than $3.5$ times greater.
In the $d=7$ case the error bound on the BB84 protocol is $D_{coh}=23.72\%$, meaning that by using error correction and privacy amplification procedures Alice and Bob may obtain a completely secure and error free shared encryption key.
However, Eve's cloning attack attempt will increase the error rate far above the bound that the BB84 protocol can handle, meaning that in those cases Alice and Bob can easily detect her and simply discard the compromised key.

In order to compare this result with the two dimensional case the researchers attempted a similar attack on a two dimensional version of the BB84 protocol (the one we described in section \ref{secQuBitBB84}) and obtained quantum bit error rates of $19 \%$ and $0.7 \%$ with and without the attempted cloning attack respectively, well above and below the security bound of the 2 dimensional BB84 protocol of $D_{coh}=11.00\%$.
This however, is a smaller bit error rate than the one incurred for a standard BB84 hack, hence, it is clear that high-dimensional quantum cryptography leads to higher signal disturbance in the presence of an optimal cloning attack, resulting in a larger tolerance to noise in the quantum channel.
\begin{figure}[H]
  \centering
  \includegraphics[scale=0.4]{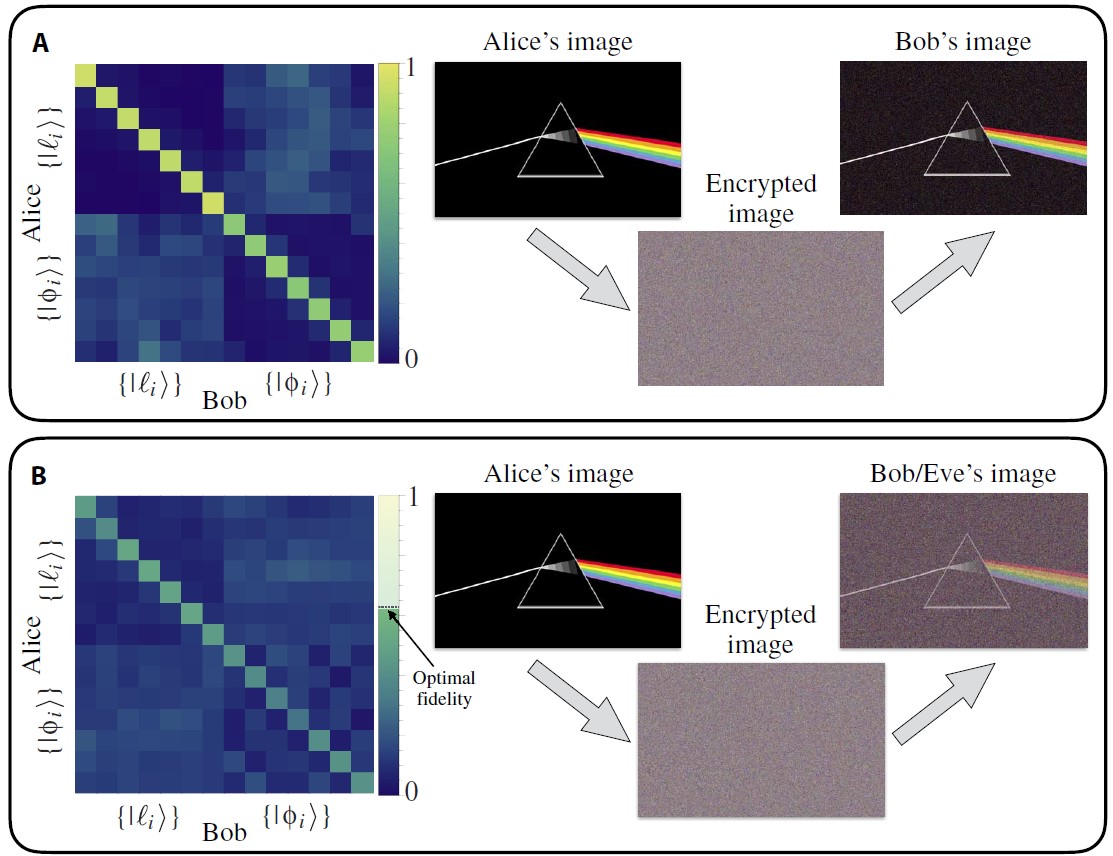}
  \caption{(Image taken from \cite{Boucharde1601915}).
  This figure is about the implementation of BB84 with and without Eavesdrop in $d=7$.
  In figure A the results without eavesdropping are shown and in sub figure B the results with eavesdropping are shown.
  On the left sides of both figures are the experimentally measured probability matrices and on the right side are Alice's messages before encryption and after Bob received and decrypted them using the shared key.}
\end{figure}

In addition to the aforementioned measurements, the researchers also tested the mutual information rate between Alice and Bob, which may be calculated form:
\begin{equation}
    I^{d}_{AB} = \log_2(d)+(1-e^N_B) \log_2(1-e^N_B)+e^N_B \log_2(\frac{e^N_B}{(d-1)})
\end{equation}
where $e^N_B$ is Bob's error rate.
Experimental values of 0.36 and 1.73 bits per photon were obtained for Alice and Bob’s mutual information with and without Eve's attempted cloning attack respectively.

The results presented in this paper illustrated the feasibility of high dimensional optimal quantum cloning of orbital angular momentum states of single photons.
This capacity was then put to use to test the venerability of the BB84 protocol to cloning based attacks, which demonstrated the robustness of high dimensional quantum cryptography upon quantum hacking.

\section{Practical implementation and resolution}
\begin{figure}[H]
  \centering
  \includegraphics[width=1\textwidth]{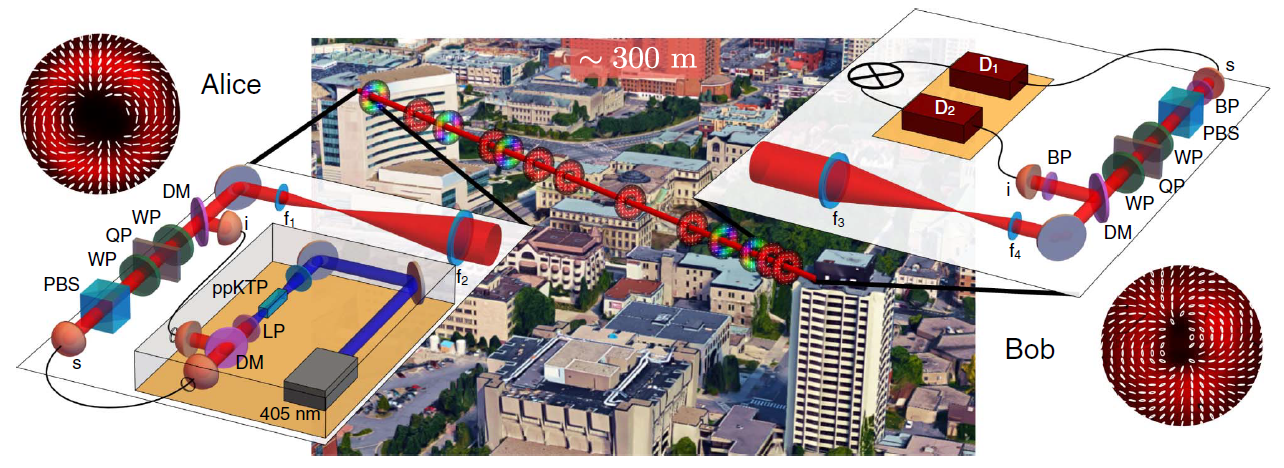}
  \caption{(Image taken from \cite{Sit17}).
  Schematic representation of the experimental setup, detailed descriptions of Alice and Bob's setups are given in sections \ref{subsubsecAlicesetup} and \ref{subsubsecBobsetup} respectively.}
  \label{figtwobuildingsconnectedbyalaser}
\end{figure}

In this section we describe in detail one particular experimental implementation of qudit based communications as an example, in particular we examine the one described in the paper \cite{Sit17}, where the researches used the BB84 protocol is 4 dimensions.
The researches who wrote the paper in question built a free-space link between the rooftops of two building $0.3$ km apart and $40$ meters above the ground, in the University of Ottawa campus, a schematic description of the system overlaid over an image of the two buildings is given in figure \ref{figtwobuildingsconnectedbyalaser}.

\subsection{The experiment setup}
\subsubsection{Alice setup}
\label{subsubsecAlicesetup}
Alice's system consists of a heralded single-photon source as well as a ppKTP crystal, two dichronic mirrors (DM), a polarizing beam splitter (PBS), two wave plates (WP), a q-plate (QP), and a telescope build from a pair of lenses whose focal lengths are $f_1 = 75$mm and $f_2 = 400$mm (items are listed in order of application as seen in figure \ref{figtwobuildingsconnectedbyalaser}, the specific number and order of WP and QP depends on the MUB being encoded).
The photons emitted by the laser become pairs of photons due to the spontaneous parametric down conversion process in the ppKTP crystal which is pumped by a laser diode.
The idler photon and the signal photon are each given a different wavelength and the two wavelengths are chosen to be non-degenerate in order to better separate the two; only the signal photon is encoded with information.
It should be noted that because of this, the scheme is still immune to number-splitting attacks, since the idler photon does not contain any of the polarization or OAM information of the signal photon.
The signal and idler photons are each coupled to a separate single-mode fiber (SMF) to spatially filter the photons into the fundamental mode.
Band-pass filters are placed in front of the fiber couplers in order to select for the correct photon pairs. 
Using a combination of wave plates and q-plates Alice then prepares the signal photon into on of the different MUBs, the specific combination and order of the plates depends on the chosen MUB Alice wishes to prepare the photon in.
The idler and signal photon are then recombined using another dichroic mirror and passed trough the telescope in order to minimize divergence upon propagation on their way before they are sent to Bob.

\subsubsection{Bob setup}
\label{subsubsecBobsetup}
Bob's system consists of a telescope made from a pair of lenses whose focal lengths are $f_3 = 400$mm and $f_4 = 50$mm, a DM, two WP, a QP, a PBS, and a pair of single photon detectors (items are listed in order of application as seen in figure \ref{figtwobuildingsconnectedbyalaser}).
In order to measure to received photons Bob first demagnifies their structure using the telescope and then he uses a DM to separate the idler photon and the signal photon.
The idler photon is the coupled into a SMF to act as a herald for the signal photon.
Meanwhile, the signal photon goes trough a series of WP and QP abd PBS which are a mirrored copy of the set Alice made.
Once this step is done Bob can Bob can make a measurement on the signal photon by projecting it onto one of the states from one of the MUBs.
By mirroring the system Alice prepared to set up her signal photon Bob has a spatial mode filter which will make it so that if he projects it onto the same state that Alice prepared it in the signal photon will be phase-flattened and optimally detected.
By using APDs and a coincidence logic box (marked in the diagram by $\otimes$), the received idler photon acts as a trigger for the arrival of the signal photon and the coincidence rates are recorded.

\subsection{Results}
\begin{figure}[H]
  \centering
  \includegraphics[width=1\textwidth]{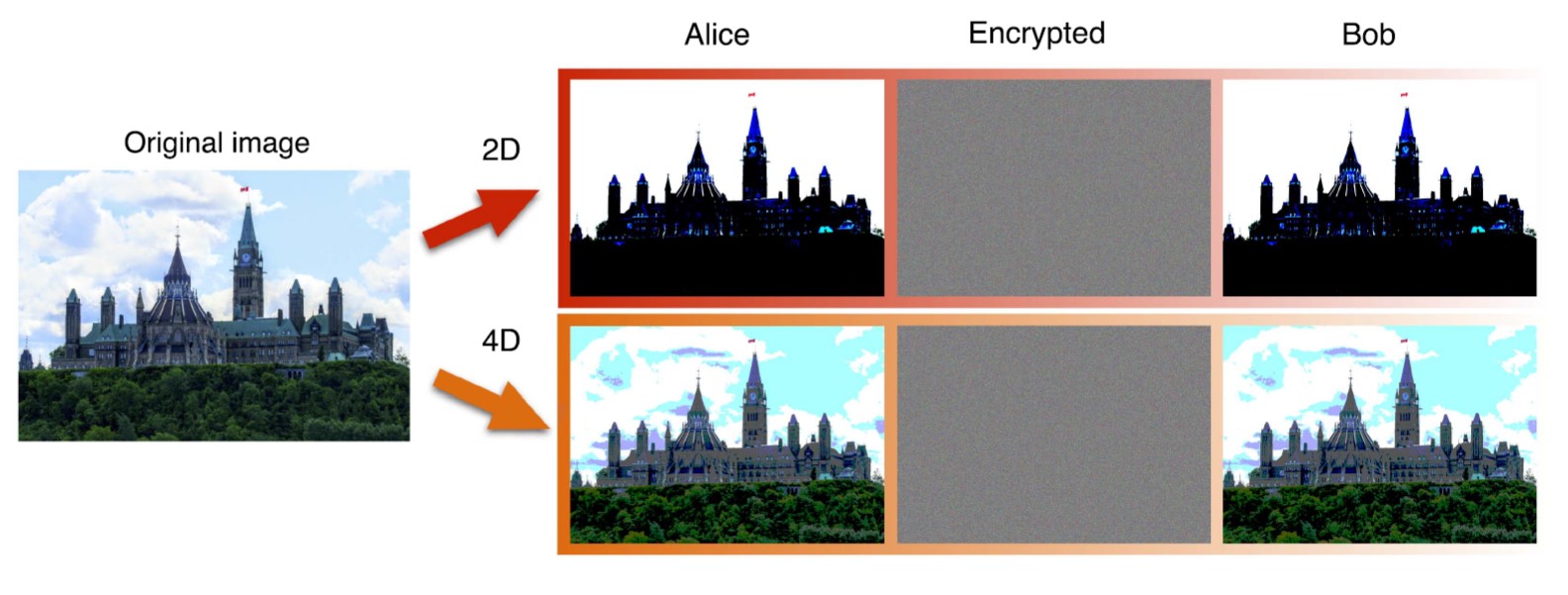}
  \caption{(Image taken from \cite{Sit17}).
  A visual illustration of the difference in key rates for $d=4$ and $d=2$. We can see that in the higher dimension the higher key rate allows more information to be transferred.}
  \label{fig4and2}
\end{figure}

As previously discussed, the key rate is defined as:
\begin{equation}
    R(e_b)=log_2(d)-2h(e_b)
\end{equation}
where $d$ is the dimension of Hilbert space, $h(e_b)$ is the Shannon entropy in dimension $d$, and $e_b$ is the bit error rate.
The theoretical maximum bit error rates for the BB84 protocol are $e_b^{d=2 \text{ (max)}} = 11 \%$ and $e_b^{d=4 \text{ (max)}} = 18.9 \%$ in two and four dimensions respectively, and the theoretical maximum key rates are $R(0) = 1$ and $R(0) = 2$ for two and four dimensions respectively.
In this experiment the values obtained in 4d were $e_b^{4=2} = 14\%$ and $R^{4d} = 0.39$, these values are sufficient for a secure key to be transmitted even without the use of any corrections.
For comparison they also performed the experiment with qubits and obtained $e_b^{d=2} = 5\%$ and $R^{2d} = 0.43$, the difference between the performance of the protocol in these two different dimensions is illustrated in figure \ref{fig4and2}

It should be noted that unlike in the experiments detailed in \cite{Bouchard2018experimental}, the experiments detailed in this paper \cite{Sit17} were performed with an open air channel over a large distance, which greatly affected the resulting bit error rate.

\section{Comparison between qubit and qudit based protocols}
\label{sectionFinalComparison}
When comparing the performance of different protocols there are five main subjects we should consider:
\begin{itemize}
  \item The bit error rate tolerance, which is the ability of the protocol to continue working as intended after a percentage of the bits were unintentionally flipped.
  This is the primary indicator of the protocol's viability for use in noisy quantum channels.
  The maximum theoretical bit rate can be computed in theory, and also measured in practice in experimental systems.
  \item The key rate, defined as the number of bits of secret key established divided by the number of sifted photons.
  \item The sifting rate, this indicates what portion of the bits that were successfully transferred from Alice to Bob will be used in the actual encryption key that the two of them will use.
  This fraction is determined by the design of the protocol.
  \item The effective key rate, which is given by the product of the key rate and the sifting rate, this number indicates what portion of each qudit of information Alice sends to Bob will become a part of the final encryption key.
  \item The difficulty and cost of implementing the hardware and software required for the protocol.
  Since at this moment all present implementations of these protocols are experimental models made in laboratory settings there is not clear way to ascertain the relative commercial viability of large scale implementations of these protocols, meaning that there is not much for us to go on in writing such a comparison at the moment.
  That said, this point nevertheless will have a large impact on what communication protocols will actually see large scale use in the future.
\end{itemize}

\begin{table}[H]
\centering
\begin{tabularx}{1.07\textwidth} { 
   >{\centering\arraybackslash}c
   >{\centering\arraybackslash}c
   >{\centering\arraybackslash}c
   >{\centering\arraybackslash}c
   >{\centering\arraybackslash}c
   >{\centering\arraybackslash}c
   >{\centering\arraybackslash}c
   >{\raggedright\arraybackslash}X  }
\hline
\hline
Protocol & $d$ & $e_b^{\text{max}}$ & $e_b^{\text{exp}}$ & $R(0)$ & $R^{\text{exp}}$ & Sifting & $R^{\text{exp}}$ x Sifting \\ 
\hline
Chau02 \cite{chau02} & $2$ & $27.6 \%$ & - & $1$ & - & $1/2$ & - \\
Chau15 \cite{Bouchard2018experimental} & $4$ & $50 \%$ & $0.778 \%$ & $1$ & $0.8170$ & $1/6$ & $0.1362$ \\
       & $8$ & $50 \%$ & $3.11 \%$ & $1$ & $0.8172$ & $1/28$ & $0.0292$ \\
BB84 \cite{Bouchard2018experimental}   & $2$ & $11.00 \%$ & $0.628 \%$ & $1$ & $0.8901$ & $1/2-1*$ & $0.4451 - 0.8901$ \\
       & $4$ & $18.93 \%$ & $3.51 \%$ & $2$ & $1.4500$ & $1/2-1*$ & $0.7250 - 1.4500$ \\
       & $8$ & $24.70 \%$ & $10.9 \%$ & $3$ &$1.3942$ & $1/2-1*$ & $0.6971 - 1.3942$ \\
BB84 \cite{Sit17} & $2$ & $11.00\%$ & $5.0\%$ & $1$ & $0.43$ & $1/2-1*$ & $0.215-0.43$ \\
       & $4$ & $18.93\%$ & $14\%$ & $2$ & $0.39$ & $1/2-1*$ & $0.195-0.39$ \\
MUB \cite{Bouchard2018experimental} & $2$ & $12.62 \%$ & $0.923 \%$ & $1$ & $0.8727$ & $1/3-1*$ & $0.2909-0.8727$ \\
       & $4$ & $23.17 \%$ & $3.87 \%$ & $2$ & $1.5316$ & $1/5-1*$ & $0.3063-1.5316$ \\
\hline
\hline
\end{tabularx}
\caption{
The data for this table is taken from  \cite{Bouchard2018experimental, chau02,chau15OGpaper, Sit17}.
The theoretical maximum tolerance for the QBER $e_b^{\text{max}}$ and the theoretical maximum key rate $R(0)$ are presented alongside their experimentally measured values $e_b^{\text{exp}}$ and $R^{\text{exp}}$.
*In the BB84 and MUB protocols depending on the size of the encryption key Alice wishes to send to Bob the choice of basis can be biased to get a a larger sifting efficiency than $1/2$ and $1/(d+1)$ for BB84 and MUB, respectively \cite{siftingratehintg}.
In the infinite key limit, the sifting efficiency can be made to approach 1 for these protocols.}
\end{table}

We find that the use of qudits in place of qubits increases error resilience and should expect it to also increase the key rate, however the exact effects differ by protocol.
\\
For the Chau15 protocol, we obtain a significant increase in the error tolerance of the protocol ($27.6\%$ to $50\%$).
This great error resistance would make it suitable to noisy channels, but would make it unwise to implement in comparatively low noise channels since in such cases all that increasing the error tolerance would do is give Eve a better window to attempt to intercept Alice's messages without being noticed, although this is somewhat counteracted by the high sifting rate.
While increasing the dimension of the qudit used in Chau15 has no effect on the key rate, it does have a significant effect on the sifting rate which equals $2/(d^2-d)$, meaning that for high dimensions the effective key rate can become so small so as to make usage of it impractically slow, such as the effective key rate of $0.0292$ found for Chau15 in 8 dimensions (assuming that the experimental key rate for $d=16$ would also be $R^{\text{exp}} = 0.817$, we would get an effective key rate of $0.0068$, requiring an average of $147$ qudits per encryption key bit).
Overall, the Chau15 protocol seems suitable for use in high noise environments and at a relatively low dimension, unless eavesdropping is a major concern in which case the high sifting rate might be preferable.
\\
The BB84 protocol appears to benefit to most from the use of qudits out of the protocols we've examined, since the error resilience and key rate both increase as we increase the qudit dimension without affecting the sifting rate.
We also found in \cite{Sit17} that the protocol remains reliable even in noisy environments (open space propagation across a relatively large distance).
Overall, BB84 seems fairly versatile in applicability for different environments, but best suited for relatively low noise environments (compared to Chau15).
In such environments we would be able to use a high dimensional qudit to fully benefit from the increased key rate, however this would also create a greater risk of eavesdropping, so the dimension of the qudit should be chosen carefully the balance these conflicting concerns.
\\
MUB-based protocols offer a middle ground between the two.
Their error resilience benefits more from the increased dimension than BB84 but not as much as Chau15.
Their (Theoretical) key rate increase as $\log_2 (d)$ but the sifting rate increases as $1/(d+1)$, so while the effective key rate doesn't decay as quickly as in Chau15, it does decay, unlike in BB84, though this can be counteracted by biasing the choice of basis \cite{siftingratehintg}.
\\
Overall, we think that in theory, difficulty of implementation not withstanding, qudits are preferable over qubits for use in QKD.
However, due to the difficulty of implantation in practice, qubits may be more reliable for real world applications.

\section{Conclusion}
\label{sectionConclusion}
In our work we have examined several different QKD protocol examining both qubit and qudit based implementations and in particular we were intereseted in comparing the benefits and drawbacks of each of them.
We found across the different protocols that we've examined that qudit based protocols seem to be preferable, given that a protocol which is appropriate to the operational environment is chosen.
However, this apparent benefit does not account for the increased difficulty of implementation, which may make qubit based protocols preferable in spite of their lesser performance in these regards.

\printbibliography

\end{document}